\documentclass[10pt]{article}
\usepackage{latexsym,graphicx}
\newcommand{\be}{\begin{equation}}
\newcommand{\ee}{\end{equation}}
\def\n{\noindent}
\catcode `\@=11 \catcode `\@=12
\begin{document}
\begin{center}
\large{\bf {TILTED BIANCHI TYPE V BULK VISCOUS COSMOLOGICAL
MODELS IN GENERAL RELATIVITY }} \\
\vspace{10mm}
\normalsize{ANIRUDH PRADHAN\footnote{Corresponding author}} \\
\vspace{5mm} \normalsize{\it{Department of Mathematics, Hindu
Post-graduate College,
 Zamania-232 331, Ghazipur, India}} \\
\normalsize{\it{email: pradhan@iucaa.ernet.in, acpradhan@yahoo.com}}\\
\vspace{5mm}
\normalsize{SUDHIR KUMAR SRIVASTAVA} \\
\vspace{5mm} \normalsize{\it{Department of Mathematics, J.J. Inter
College, Gandhi Nagar-233 225, Ghazipur, India}} \\
\end{center}
\vspace{10mm}
\begin{abstract}
Conformally flat tilted Bianchi type V cosmological models in
presence of a bulk viscous fluid and heat flow are investigated.
The coefficient of bulk viscosity is assumed to be a power
function of mass density. The cosmological constant is found to
be a decreasing function of time, which is supported by results
from recent type Ia supernovae observations. Some physical and
geometric aspects of the models are also discussed.
\end{abstract}
\smallskip
\n Key words : cosmology, tilted Bianchi type V universe, variable cosmological constant\\
\n PACS: 98.80.Jk, 98.80.Es\\
\newpage
\section{INTRODUCTION}
The Bianchi cosmologies play an important role in theoretical cosmology and
have been much studied since the 1960s. A Bianchi cosmology represents a spatially
homogeneous universe, since by definition the spacetime admits a three-parameter
group of isometries whose orbits are spacelike hyper-surfaces. These models can be
used to analyze aspects of the physical Universe which pertain to or which may be
affected by anisotropy in the rate of expansion, for example , the cosmic microwave
background radiation, nucleosynthesis in the early universe, and the question of the
isotropization of the universe itself \cite{ref1}. Spatially homogeneous cosmologies
also play an important role in attempts to understand the structure and properties
of the space of all cosmological solutions of Einstein field equations. A spatially
homogeneous cosmology is said to be {\it tilted} if the fluid velocity vector is not
orthogonal to the group orbits, otherwise the model is said to be {\it non-tilted}
\cite{ref2}. A tilted model is spatially homogeneous relative to observers whose
world line are orthogonal relative to the group orbits, but is spatially inhomogeneous
relative to observers comoving with the fluid. In a tilted Bianchi cosmology the tilt
can become extreme in a finite time as measured along the fluid congruence, with the
result that the group orbits become time-like. This means that the models are no longer
spatially homogeneous \cite{ref3}. \\

The general dynamics of tilted models have been studied by King
and Ellis \cite{ref2}, and Ellis and King \cite{ref4}. Ellis and
Baldwin \cite{ref5}  have shown that we are likely to be living
in a tilted universe and they have indicated how we may detect it.
Beesham \cite{ref6} derived tilted Bianchi type V cosmological
models in the scale-covariant theory. A tilted cold dark matter
cosmological scenario has been discussed by Cen et al.
\cite{ref7}. Several researchers (Matravers et al. \cite{ref8},
Ftaclas and Cohen \cite{ref9}, Hewitt and Wainwright
\cite{ref10}, Lidsey \cite{ref11}, Bali and Sharma \cite{ref12},
Hewitt et al. \cite{ref13}, Horwood et al. \cite{ref14},
Bogoyavlenskii and Novikov \cite{ref15}, Barrow and Sonoda
\cite{ref16}, Barrow and Hervik \cite{ref17}, Apostolopoulos
\cite{ref18}, Pradhan and Rai \cite{ref19} and Hervik
\cite{ref20} have studied various aspects of
tilted cosmological models. \\

A considerable interest has been shown to the study of physical properties of
spacetimes which are conformal to certain well known gravitational fields. The
general theory of relativity is believed by a number of unknown functions - the
ten components of $g^{ij}$. Hence there is a little hope of finding physically
interesting results without making reduction in their number. In conformally
flat spacetime the number of unknown functions is reduced to one. The
conformally flat metrices are of particular interest in view of their
degeneracy in the context of Petrov classification. A number of conformally
flat physically significant spacetimes are known like Schwarzschild interior
solution and Lema\^{i}tre cosmological universe.\\

General Relativity describes the state in which radiation concentrates around a star.
Klein \cite{ref21} worked on it and obtained an approximate solution to Einsteinian
field equations in spherical symmetry for a distribution of diffused radiation. Many
other researchers (Singh and Sattar \cite{ref22}, Roy and Bali \cite{ref23} ) have
worked on this topic and  obtained exact static spherically and cylindrically symmetric
solutions of Einstein's field equations with exception as well. Roy and Singh \cite{ref24}
have obtained a non-static plane symmetric spacetime filled with disordered radiation.
Teixeira, Wolk and Som \cite{ref25} investigated a model filled with source free
disordered distribution of electromagnetic radiation in Einstein's general relativity.
The cosmological models with heat flow have been also studied by Coley and Tupper
\cite{ref26}, Roy and Banerjee \cite{ref27}. Recently Bali and Meena \cite{ref28} have
investigated two tilted cosmological models filled with disordered radiation of perfect
fluid and heat flow. \\

Most cosmological models assume that the matter in the universe
can be described by `dust'(a pressure-less distribution) or at
best a perfect fluid. Nevertheless, there is good reason to
believe that - at least at the early stages of the universe -
viscous effects do play a role (Israel and Vardalas \cite{ref29},
Klimek \cite{ref30}, Weinberge \cite{ref31}). For example, the
existence of the bulk viscosity is equivalent to slow process of
restoring equilibrium states (Landau and Lifshitz \cite{ref32}).
The observed physical phenomena such as the large entropy per
baryon and remarkable degree of isotropy of the cosmic microwave
background radiation suggest analysis of dissipative effects in
cosmology. Bulk viscosity is associated with the GUT phase
transition and string creation. Thus, we should consider the
presence of a material distribution other than a perfect fluid to
have realistic cosmological models (see Gr\o n \cite{ref33} ) for
a review on cosmological models with bulk viscosity). The model
studied by Murphy \cite{ref34} possessed an interesting feature
in that the big bang type of singularity of infinite spacetime
curvature does not occur to be a finite past. However, the
relationship assumed by Murphy between the viscosity coefficient
and the matter density is not acceptable at large density. The
effect of bulk viscosity on the cosmological evolution has been
investigated by a number of authors in the framework of general
theory of relativity (Pavon \cite{ref35}, Padmanabhan and Chitre
\cite{ref36}, Johri and Sudarshan \cite{ref37}, Maartens
\cite{ref38}, Zimdahl \cite{ref39}, Santos {\it et al.}
\cite{ref40}, Pradhan, Sarayakar and Beesham \cite{ref41},
Kalyani and Singh \cite{ref42}, Singh, Beesham and Mbokazi
\cite{ref43}, Pradhan et al. \cite{ref44}). This motivates
to study cosmological bulk viscous fluid model.\\

Models with a dynamic cosmological term $\Lambda (t)$ are
becoming popular as they solve the cosmological constant problem
in a natural way. There is significant observational evidence for
the detection of Einstein's cosmological constant, $\Lambda$ or a
component of material content of the universe that varies slowly
with time and space and so acts like $\Lambda$. Recent
cosmological observations by High -z Supernova Team and Supernova
Cosmological Project (Garnavich et al. \cite{ref45}, Perlmutter
et al. \cite{ref46}, Riess et al. \cite{ref47}, Schmidt et al.
\cite{ref48}) suggest the existence of a positive cosmological
constant $\Lambda$ with magnitude $\Lambda (G \hbar /c^{3})
\approx 10^{-123}$. These observations on magnitude and red-shift
of type Ia supernova suggest that our universe may be a
accelerating with a large function of the cosmological density in
the form of the cosmological $\Lambda$-term. Earlier researchers
on this topic, are contained in Zeldovich \cite{ref49}, Weinberg
\cite{ref50}, Dolgov \cite{ref51}, Bertolami \cite{ref52}, Ratra
and Peebles \cite{ref53}, Carroll, Press and Turner \cite{ref54}.
Some of the recent discussions on the cosmological constant
``problem'' and consequence on cosmology with a time-varying
cosmological constant have been discussed by Dolgov \cite{ref55},
Tsagas and Maartens \cite{ref56}, Sahni and Starobinsky
\cite{ref57}, Peebles \cite{ref58}, Padmanabhan \cite{ref59},
Vishwakarma \cite{ref60}, and Pradhan et al. \cite{ref61}. This
motivates us to study the cosmological models in which
$\Lambda$ varies with time.\\

Recently Bali and Meena \cite{ref62} have investigated two
conformally flat tilted Bianchi type V cosmological models filled
with a perfect fluid and heat conduction. Conformally flat tilted
Bianchi type V cosmological models in presence of a bulk viscous
fluid and heat flow are investigated by Pradhan and Rai
\cite{ref63}. In this paper, we propose to find tilted Bianchi
type V cosmological models in presence of a bulk viscous fluid
and heat flow with variable cosmological term $\Lambda$ and we
will generalize the solutions of Refs. \cite{ref62,ref63}. This
paper is organized as follows. The metric and field equations are
presented in Section $2$. In Section $3$, we deal with the
solutions of the field equations in presence of bulk viscous
fluid. In Section $4$, we give the concluding remarks.

\section{THE METRIC AND FIELD  EQUATIONS}
We consider the Bianchi type V metric in the form
\begin{equation}
\label{eq1}
ds^{2} = - dt^{2} + A^{2} dx^{2} + B^{2} e^{2x} \left(dy^{2} + dz^{2}\right),
\end{equation}
where A, B are function of $t$ only.\\
The Einstein's field equations (in gravitational units $c = 1$, $G = 1$) read as
\begin{equation}
\label{eq2}
R^{j}_{i} - \frac{1}{2} R g^{j}_{i} + \Lambda g^{j}_{i} = -8\pi T^{j}_{i},
\end{equation}
where $R^{j}_{i}$ is the Ricci tensor; $R$ = $g^{ij} R_{ij}$ is the
Ricci scalar; $\Lambda$ is the variable cosmological constant and $T^{j}_{i}$ is the
stress energy-tensor in the presence of bulk stress given by
\begin{equation}
\label{eq3}
T^{j}_{i} = (\rho + \bar{p})v_{i}v^{j} + \bar{p} g^{j}_{i} +
q_{i}v^{j} + v_{i}q^{j},
\end{equation}
and
\begin{equation}
\label{eq4}
\bar{p} = p - \xi v^{i}_{;i}.
\end{equation}
Here $\rho$, $p$, $\bar{p}$ and $\xi$ are the energy density,
isotropic pressure, effective pressure and  bulk viscous
coefficient respectively and $v_{i}$ is the flow vector satisfying
the relations
\begin{equation}
\label{eq5}
g_{ij} v^{i}v^{j} = - 1,
\end{equation}
\begin{equation}
\label{eq6}
q_{i} q^{j} > 0,
\end{equation}
\begin{equation}
\label{eq7}
q_{i}v^{i} = 0,
\end{equation}
where $q_{i}$ is the  heat conduction vector orthogonal to $v_{i}$.
The fluid flow vector has the components $(\frac{\sinh \lambda}{A}, 0, 0,
\cosh \lambda)$ satisfying Eq. (\ref{eq5}) and $\lambda$ is the tilt angle.\\
The Einstein's field equations (\ref{eq2}) for the line element (\ref{eq1})
has been set up as
\begin{equation}
\label{eq8}
- 8\pi\left[(\rho + \bar{p})\sinh^{2}  \lambda + \bar{p} + 2 A q_{1} \sinh \lambda \right]
= \frac{2 B_{44}}{B} + \left(\frac{B_{4}}{B}\right)^{2} - \frac{1}{A^{2}} - \Lambda,
\end{equation}
\begin{equation}
\label{eq9}
- 8\pi \bar{p} = \frac{A_{44}}{A} + \frac{B_{44}}{B} + \frac{A_{4}B_{4}}{AB}
- \frac{1}{A^{2}} - \Lambda,
\end{equation}
\begin{equation}
\label{eq10}
- 8\pi\left[- (\rho + \bar{p})\cosh^{2} \lambda + \bar{p} - 2 A q_{1} \sinh \lambda \right]
 = \frac{2 A_{4}B_{4}}{AB} + \left(\frac{B_{4}}{B}\right)^{2} - \frac{3}{A^{2}} - \Lambda,
\end{equation}
\begin{equation}
\label{eq11}
-8 \pi\left[(\rho + \bar{p})A \sinh \lambda ~  \cosh \lambda + A^{2} q_{1}
(\cosh \lambda + \sinh \lambda ~ \tanh \lambda)\right] = \frac{2 A_{4}}{A}
- \frac{2 B_{4}}{B} - \Lambda,
\end{equation}
where the suffix $4$ at the symbols $A$, $B$ denotes ordinary
differentiation with respect to $t$.
\section{SOLUTION OF THE FIELD EQUATIONS}
Equations (\ref{eq8}) - (\ref{eq11}) are four independent
equations in eight unknowns $A$, $B$, $\rho$, $p$, $\xi$, $q$,
$\Lambda$ and $\lambda$. For the complete
determinacy of the system, we need four extra conditions.\\
First we assume that the spacetime is conformally flat which leads to
\begin{equation}
\label{eq12}
C_{2323} = \frac{1}{3}\left[\frac{A_{44}}{A} - \frac{B_{44}}{B} - \frac{A_{4}B_{4}}
{AB} + \frac{B^{2}_{4}}{B^{2}}\right] = 0
\end{equation}
and secondly, we assume
\begin{equation}
\label{eq13}
A = B^{n},
\end{equation}
where $n$ is any  real number.
Eqs. (\ref{eq12}) and (\ref{eq13}) lead to
\begin{equation}
\label{eq14}
\frac{B_{44}}{B} + (n - 1)\frac{B^{2}_{4}}{B^{2}} = 0.
\end{equation}
From  Equations (\ref{eq8}), (\ref{eq10}) and (\ref{eq13}), we have
\begin{equation}
\label{eq15}
- 4 \pi \left[(\rho + \bar{p}) \cosh 2\lambda + 4 B^{n} q_{1} \sinh \lambda\right]
= \frac{B_{44}}{B} - n \frac{B^{2}_{4}}{B^{2}} + \frac{1}{B^{2n}},
\end{equation}
and
\begin{equation}
\label{eq16}
4 \pi (\rho - \bar{p}) = \frac{B_{44}}{B} + (n + 1) \frac{B^{2}_{4}}{B^{2}}
- \frac{2}{B^{2n}} - \Lambda.
\end{equation}
Equations (\ref{eq9}), (\ref{eq12})  and (\ref{eq16}) lead to
\begin{equation}
\label{eq17}
-n(2n - 3)\frac{B^{2}_{4}}{B^{2}} - (2n - 1) \frac{B_{44}}{B} -
\frac{1}{B^{2n}} = 4 \pi (\rho + \bar{p}).
\end{equation}
Equations (\ref{eq11}) and (\ref{eq13}) lead to
\begin{equation}
\label{eq18}
- 16 \pi q_{1} B^{n} \sinh \lambda = \frac{[2(n - 1)B_{4} - \Lambda B] \tanh 2\lambda}
{B^{n + 1}} + 4\pi (\rho + \bar{p})\sinh 2\lambda~ \tanh 2\lambda.
\end{equation}
From Eqs. (\ref{eq15}) and (\ref{eq18}), we obtain
\begin{equation}
\label{eq19}
\frac{B_{44}}{B} - \frac{n B^{2}_{4}}{B^{2}} + \frac{1}{B^{2n}} =
- \frac{4\pi(\rho + \bar{p})}{\cosh 2\lambda} +
\frac{[2(n - 1) B_{4} - \Lambda B] \tanh 2\lambda}{B^{n + 1}}.
\end{equation}
Equations (\ref{eq17}) and (\ref{eq19}) lead to
\[
\frac{B_{44}}{B} - \frac{n B^{2}_{4}}{B^{2}} + \frac{1}{B^{2n}} = \frac{1}{\cosh 2\lambda}
\left[n(2n - 3)\frac{B^{2}_{4}}{B^{2}} + (2n - 1)\frac{B_{44}}{B} + \frac{1}{B^{2n}}\right]
\]
\begin{equation}
\label{eq20}
+ \frac{[2(n - 1)B_{4} - \Lambda B]\tanh 2\lambda}{B^{n + 1}}.
\end{equation}
Equation (\ref{eq14}) can be rewritten as
\begin{equation}
\label{eq21}
\frac{B_{44}}{B_{4}} + \frac{(n - 1) B_{4}}{B} = 0,
\end{equation}
which on integration leads to
\begin{equation}
\label{eq22}
B = n^{\frac{1}{n}} ~ (\alpha t + \beta)^{\frac{1}{n}}.
\end{equation}
where $\alpha$, $\beta$ are constants of integration. Hence we obtain
\begin{equation}
\label{eq23}
A^{2} = n^{2} ~ (\alpha t + \beta)^{2},
\end{equation}
\begin{equation}
\label{eq24}
B^{2} = n^{\frac{2}{n}} ~ (\alpha t + \beta)^{\frac{2}{n}}.
\end{equation}
Hence the geometry of the spacetime (\ref{eq1}) reduces to the form
\begin{equation}
\label{eq25}
ds^{2} = - dt^{2} + n^{2} (\alpha t + \beta)^{2} dx^{2} + [n(\alpha t + \beta)]^
{\frac{2}{n}} e^{2x} (dy^{2} + dz^{2}).
\end{equation}
After the suitable transformation of coordinates, the metric (\ref{eq25})
takes the form
\begin{equation}
\label{eq26}
ds^{2} = - \frac{dT^{2}}{2} + n^{2} T^{2} dX^{2} + n^{\frac{2}{n}}
T^{\frac{2}{n}} e^{2X}(dY^{2} + dZ^{2}).
\end{equation}
The effective pressure and density of the model (\ref{eq26}) are given by
\begin{equation}
\label{eq27}
8\pi \bar{p} = 8\pi (p -\xi \theta) = - \frac{(\alpha^{2} - 1)}{n^{2} T^{2}} -  \Lambda,
\end{equation}
\begin{equation}
\label{eq28}
8\pi \rho = \frac{3(\alpha^{2} - 1)}{n^{2} T^{2}} + \Lambda,
\end{equation}
where $\theta$ is the scalar of expansion calculated for the flow vector
$v^{i}$ and given is as
\begin{equation}
\label{eq29}
\theta = \frac{2K + (n + 2)k\alpha}{nT}.
\end{equation}
The tilt angle $\lambda$ is given by
\begin{equation}
\label{eq30}
\cosh^{2}  \lambda = k^{2},
\end{equation}
\begin{equation}
\label{eq31}
\sinh^{2}  \lambda = K^{2},
\end{equation}
where $k$ and $K$ are constants given by
\begin{equation}
\label{eq32}
k^{2} = \frac{(n\alpha^{2} - 1)^{2}}{(\alpha^{2} - 1) (2n^{2} \alpha^{2}
- n \alpha^{2} - 1) + 2\alpha^{2} (n - 1)\sqrt{(n^{2} - n)(1 - \alpha^{2})}},
\end{equation}
\begin{equation}
\label{eq33}
K^{2} = \frac{(n - 1)(2n \alpha^{2} - n\alpha^{4} - \alpha^{2})
- 2\alpha^{2} (n - 1)\sqrt{(n^{2} - n)(1 - \alpha^{2})}}{(\alpha^{2} - 1)
(2n^{2} \alpha^{2} - n \alpha^{2} - 1) + 2\alpha^{2} (n - 1)\sqrt{(n^{2} - n)
(1 - \alpha^{2})}}.
\end{equation}
For the specification of $\xi$, we assume that the fluid obeys an equation of
state of the form
\begin{equation}
\label{eq34}
p = \gamma \rho,
\end{equation}
where $\gamma(0 \leq \gamma \leq 1)$ is a constant. \\
Thus, given $\xi(t)$ we can solve the system for the physical quantities.
Therefore to apply the third condition, let us assume the following {\it adhoc}
law \cite{ref38,ref39}
\begin{equation}
\label{eq35}
\xi(t) = \xi_{0} \rho^{m},
\end{equation}
where $\xi_{0}$ and $m$ are real constants. If $m = 1$, Eq. (\ref{eq34})
may correspond to a radiative fluid \cite{ref50}, whereas
$m$ = $\frac{3}{2}$ may correspond to a string-dominated universe. However,
more realistic models \cite{ref40} are based on lying the regime
$0 \leq m \leq \frac{1}{2}$. \\
\subsection {MODEL I: SOLUTION FOR $(\xi = \xi_{0})$.}
When $m = 0$, Eq. (\ref{eq35}) reduces to $\xi = \xi_{0}$ = constant and hence
Eq. (\ref{eq27}), with the use of (\ref{eq34}) and (\ref{eq28}), leads to
\begin{equation}
\label{eq36}
4\pi (1 + \gamma)\rho = \frac{4\pi \xi_{0}[2K + (n + 2)k\alpha]}{nT} +
\frac{(\alpha^{2} - 1)}{n^{2}T^{2}}.
\end{equation}
Eliminating $\rho(t)$ between (\ref{eq28}) and (\ref{eq36}), we get
\begin{equation}
\label{eq37}
(1 + \gamma)\Lambda = \frac{8\pi \xi_{0}[2K + (n + 2)k\alpha]}{nT} -
\frac{(\alpha^{2} - 1)(1 + 3\gamma)}{n^{2}T^{2}}.
\end{equation}
\subsection {MODEL II: SOLUTION FOR $(\xi = \xi_{0}\rho)$}
When $m = 1$, Eq. (\ref{eq35}) reduces to $\xi = \xi_{0}\rho$ and hence
Eq. (\ref{eq27}), with the use of (\ref{eq34}) and (\ref{eq28}), leads to
\begin{equation}
\label{eq38}
4\pi \rho = \frac{(\alpha^{2} - 1)}{nT[n(1 + \gamma)T - \xi_{0}\{2K + (n + 2)
k\alpha\}]}.
\end{equation}
Eliminating $\rho(t)$ between (\ref{eq28}) and (\ref{eq38}), we get
\begin{equation}
\label{eq39}
\Lambda = \frac{(\alpha^{2} - 1)}{nT}\left[\frac{2}{[n(1 + \gamma)T - \xi_{0}
\{2K + (n + 2)k\alpha\}]} - \frac{3}{nT}\right].
\end{equation}
\subsection {SOME PHYSICAL AND GEOMETRIC PROPERTIES OF THE MODELS}
From Eqs. (\ref{eq37}) and (\ref{eq39}), we observe that the
cosmological constant in both the models is a decreasing function
of time and it approaches a small and positive value for large
$T$ (i.e. the present epoch) which is supported by the results
from recent type Ia supernovae observations (Garnavich et
al.\cite{ref45}, Perlmutter et al. \cite{ref46}, Riess et al.
\cite{ref47}, Schmidt et al.\cite{ref48}). \\
The weak and strong energy conditions, we have, in Model I
\begin{equation}
\label{eq40}
\rho + p = \frac{(\alpha^{2} - 1)}{4\pi n^{2} T^{2}} + \frac{\xi_{0} \{2K +
(n + 2) k \alpha \}}{n T},
\end{equation}
\begin{equation}
\label{eq41}
\rho - p = \frac{(\alpha^{2} - 1)(1 - \gamma)}{4\pi (1 + \gamma)n^{2} T^{2}} +
\frac{(1 - \gamma)\xi_{0} \{2K + (n + 2) k \alpha \}}{(1 + \gamma)n T},
\end{equation}
\begin{equation}
\label{eq42}
\rho + 3p =  \frac{(\alpha^{2} - 1)(1 + 3\gamma)}{4\pi (1 + \gamma)n^{2} T^{2}} +
\frac{(1 + 3\gamma)\xi_{0} \{2K + (n + 2) k \alpha \}}{(1 + \gamma)n T},
\end{equation}
\begin{equation}
\label{eq43}
\rho - 3p = \frac{(\alpha^{2} - 1)(1 - 3\gamma)}{4\pi (1 + \gamma)n^{2} T^{2}} +
\frac{(1 - 3\gamma)\xi_{0} \{2K + (n + 2) k \alpha \}}{(1 + \gamma)n T}.
\end{equation}
In Model II, we have
\begin{equation}
\label{eq44}
\rho + p = \frac{(1 + \gamma)(\alpha^{2} - 1)}{4\pi nT[n(1 + \gamma)T - \xi_{0}
\{2K + (n + 2)k\alpha\}]},
\end{equation}
\begin{equation}
\label{eq45}
\rho - p = \frac{(1 - \gamma)(\alpha^{2} - 1)}{4\pi nT[n(1 + \gamma)T - \xi_{0}
\{2K + (n + 2)k\alpha\}]},
\end{equation}
\begin{equation}
\label{eq46}
\rho + 3p = \frac{(1 + 3 \gamma)(\alpha^{2} - 1)}{4\pi nT[n(1 + \gamma)T - \xi_{0}
\{2K + (n + 2)k\alpha\}]},
\end{equation}
\begin{equation}
\label{eq47}
\rho - 3p = \frac{(1 - 3\gamma)(\alpha^{2} - 1)}{4\pi nT[n(1 + \gamma)T - \xi_{0}
\{2K + (n + 2)k\alpha\}]}.
\end{equation}
The reality conditions $\rho \geq 0$, $p \geq 0$ and $\rho - 3p  \geq 0$
impose further restrictions on both of these models.\\
The flow vector $v^{i}$ and heat conduction vector $q^{i}$ for the models
(\ref{eq26}) are obtained as
\begin{equation}
\label{eq48}
v^{1} = \frac{K}{n T},
\end{equation}
\begin{equation}
\label{eq49}
v^{4} = k,
\end{equation}
\begin{equation}
\label{eq50}
q^{1} = - \frac{k \{(\alpha^{2} - 1)k  K + (n - 1)\alpha\}}
{4\pi n^{3} T^{3} (k^{2} + K^{2})},
\end{equation}
\begin{equation}
\label{eq51}
q^{4} =  - \frac{K \{(\alpha^{2} - 1)k K + (n - 1)\alpha\}}
{4\pi n^{3} T^{3} (k^{2} + K^{2})}.
\end{equation}
The rate of expansion $H_{i}$ in the direction of $X$, $Y$, $Z$-axes
are given by
\begin{equation}
\label{eq52}
H_{1} = \frac{\alpha}{T},
\end{equation}
\begin{equation}
\label{eq53}
H_{2} = H_{3} = \frac{\alpha}{n T}.
\end{equation}
The non-vanishing components of shear tensor $(\sigma_{ij})$ and rotation
tensor $(\omega_{ij})$ are obtained as
\begin{equation}
\label{eq55}
\sigma_{11} = \frac{2}{3} n k^{2} T [(n - 1)k \alpha - K],
\end{equation}
\begin{equation}
\label{eq55}
\sigma_{22} = \sigma_{33} = \frac{(nT)^{\frac{2}{n} - 1} e^{2X}}{3}
[(1 - n)k \alpha - 2K],
\end{equation}
\begin{equation}
\label{eq56}
\sigma_{44} =  \frac{2 K^{2}}{3 n T}[(n - 1)k \alpha - K],
\end{equation}
\begin{equation}
\label{eq57}
\sigma_{14} = \frac{K}{3}[2(1 - n) k^{2} + 2 k K - 3 n],
\end{equation}
\begin{equation}
\label{eq58}
\omega_{14} = n \alpha K.
\end{equation}
The models, in general, represent shearing and  rotating universes. The models
start expanding with a big bang  at $T = 0$ and the expansion
in the models decreases as time increases and the expansion in the models stops
at $T = \infty$ and $\alpha = - \frac{2K}{(n + 2) k}$. Both density and pressure
in the models become zero at $ T = \infty$. For $\alpha = 1$, $n = 1$, we observe
that heat conduction vector $q^{1} = q^{4} = 0$. When $T \rightarrow \infty$, $v^{1}
= 0$, $v^{4} = $ constant, $q^{1} = q^{4} = 0$. Since $\lim_{T\rightarrow \infty}
\frac{\sigma}{\theta} \ne 0$, the models do not approach isotropy for large values
of $T$. There is a real physical singularity in the model at $T = 0$.\\
In case $\Lambda = 0$ and $\xi_{0} = 0$, metric (\ref{eq26}) with
expressions $p$, $\rho$, $\theta$ and $\sigma$ for this model are
same as that of solution (2.27) of Bali and Meena \cite{ref62}. In
case $\Lambda = 0$, metric (\ref{eq26}) with expressions
$\bar{p}$, $\rho$, $\theta$ and $\sigma$ for this model are same
as that
of solution (56) of Pradhan and Rai \cite{ref63}. \\
\section{CONCLUSIONS}
In this paper we have described a new class of conformally flat tilted Bianchi
type V magnetized cosmological models with a bulk viscous fluid as the source
of matter. Generally, the models are expanding, shearing and rotating.
In all these models, we observe that they do not approach isotropy for
large values of time $T$ in the presence of magnetic field. It is seen that
the solutions obtained by Bali and Meena \cite{ref62} and Pradhan and Rai
\cite{ref63} are particular cases of our solutions.\\
The coefficient of bulk viscosity is assumed to be a power function of
mass density. The effect of bulk viscosity is to introduce a change in the
perfect fluid model. We also observe here that the conclusion of Murphy
\cite{ref34} about the absence of a big bang type of singularity in the finite
past in models with bulk viscous fluid is, in general, not true.\\
The cosmological constant in all models in Sections $3.1$ and
$3.2$, are decreasing function of time and they all approach a
small positive value at late time. These results are supported by
the results from recent supernovae Ia observations recently
obtained by High - Z Supernova Team and Supernova Cosmological Project [45 - 48]. \\
\section*{Acknowledgements} A. Pradhan would like to thank the Inter-University Center for
Astronomy and Astrophysics, Pune, India for providing facility
where part of this work was carried out.
\newline

\end{document}